\documentclass[a4paper,fleqn]{cas-dc}
\usepackage[numbers, authoryear]{natbib}
\usepackage{siunitx}
\usepackage{amsmath} % à mettre dans le préambule
\usepackage{float}
\usepackage{graphicx} % souvent nécessaire pour float/table
\usepackage{placeins}     % pour \FloatBarrier si besoin
\usepackage{pdflscape} % à mettre dans le préambule
\usepackage{makecell}
\usepackage{geometry}
\usepackage{rotating} % à mettre dans le préambule
\usepackage[pagewise,switch,mathlines]{lineno}

\definecolor{darkgreen}{RGB}{0,100,0} 
\definecolor{darkred}{RGB}{150,0,0} 

\def\tsc#1{\csdef{#1}{\textsc{\lowercase{#1}}\xspace}}
\tsc{WGM}
\tsc{QE}
\begin{document}
\let\WriteBookmarks\relax
\def\floatpagepagefraction{1}
\def\textpagefraction{.001}
%\linenumbers

\shorttitle{}    

\shortauthors{A.Leclef et al}  

\title [mode = title]{Properties of Seasonal Ice at Sisyphi Cavi and Implications for Current Modification of Martian Gullies.}  


\tnotetext[1]{}

\author[1]{A. Leclef }[type=editor,
       orcid=0009-0008-0423-9525]%[<options>]

\ead{apolline.leclef@universite-paris-saclay.fr}

\affiliation[1]{organization={Institut d'Astrophysique Spatiale},
            addressline={Université Paris-Saclay}, 
            city={Orsay},
            postcode={91405 }, 
            country={France}}

\author[1]{M. Vincendon }%[]
\author[1]{C. Lantz }%[]
\author[2]{F. Andrieu }%[]
\author[1,3]{M. Ausseresse}%[]
\author[1,4]{J. Carter }%[]
\author[5]{S. J. Conway }%[]
\author[5]{M. Massé }%[]
\author[5]{K. Pasquon }%[]
\author[2,6]{F. Schmidt }%[]

\affiliation[2]{organization={Université Paris-Saclay},
            addressline={CNRS, GEOPS}, 
            city={Orsay},
            postcode={91405}, 
            country={France}}

\affiliation[3]{organization={Dept. d’Astrophysique/AIM, CEA/IRFU, CNRS/INSU},
            addressline={Université Paris et Paris-Saclay}, 
            city={Gif-sur-Yvette},
            postcode={91191}, 
            country={France}}

\affiliation[4]{organization={Laboratoire d'Astrophysique de Marseille},
            addressline={Aix Marseille Université}, 
            city={Marseille},
            postcode={13388}, 
            country={France}}

\affiliation[5]{organization={Laboratoire de Panétologie et Géosciences},
            addressline={Université de Nantes}, 
            city={Nantes},
            postcode={44109}, 
            country={France}}

\affiliation[6]{organization={Institut Universitaire de France},
            addressline={IUF}, 
            city={Paris},
            postcode={75231}, 
            country={France}}

\begin{abstract} 
Martian gullies are geologically recent landforms that may form either through liquid-water activity and/or through processes involving CO\textsubscript{2} ice. We investigated the mechanisms responsible for the formation or modification of these features by focusing on the active site of Sisyphi Cavi (68°S, 1°E), located outside the typical latitude range of gully presence. Using CRISM and OMEGA infrared data, we characterized the composition and physical state of seasonal surface ices to test the relevance of H\textsubscript{2}O and CO\textsubscript{2} ice driven mechanisms. Our analysis shows that H\textsubscript{2}O ice is not detected as an independent surface deposit, although it may be present as minor inclusions within the CO\textsubscript{2} ice layer. In particular, during the final phase of CO\textsubscript{2} ice sublimation in late spring, no H\textsubscript{2}O ice signature is observed. In the area, faint spectral signatures of sulfate salts are observed, but their distribution and amount do not suggest any direct link with gully activity.  Available observations during early and mid-spring reveal that CO\textsubscript{2} ice is translucent during these times, suggesting that it likely remains in this state through most of the ice season. However, a temporal mismatch between dark spot formation - indicative of CO\textsubscript{2} geysers through translucent ice - and gully modification (respectively occurring late winter to early spring, and mid to late spring) exists. This does not suggest a systematic link between both processes.
Overall, the available observations provide no evidence that liquid water contributes to present-day gully activity at Sisyphi Cavi, while offering no support either for the hypothesis that gully modifications are mainly driven by the formation of CO\textsubscript{2} geysers. Gully modifications at Sisyphi Cavi, observed during the late sublimation stages of CO\textsubscript{2} ice, may be rather more appropriately explained by CO\textsubscript{2}-ice-based fluidization or avalanche processes.
\end{abstract}

%\item 
%\item 
%\item 

% Keywords

\maketitle

% Main text
\section{Introduction}\

\begin{figure*}[htbp]
    \centering
    \includegraphics[width=\textwidth]{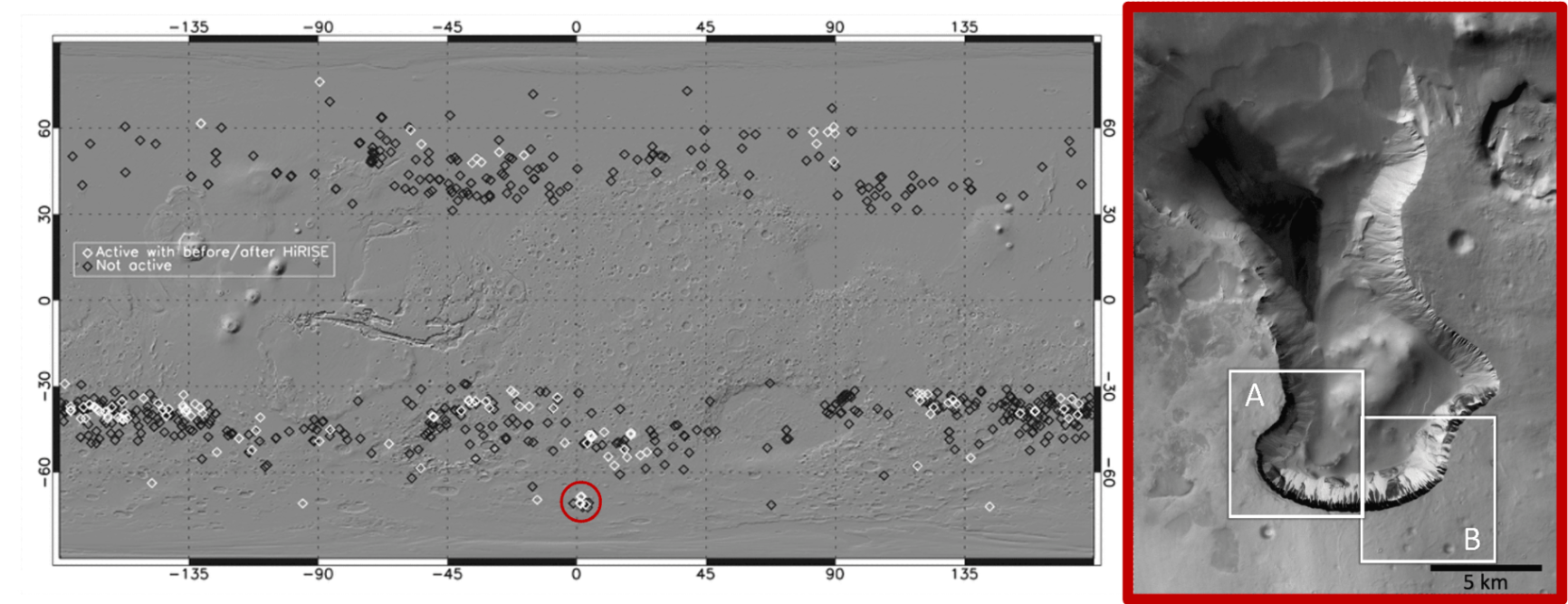}
    \caption{\textbf{The Sisyphi Cavi area analysed in this study.} Left panel, gullies map extracted from \cite{Dundas2022}, highlighting active gullies (in white) and providing overall context. Sisyphi Cavi, centred on 68°S 1°E is outlined in red. Right panel, high-resolution view of the specific sector of this region analysed in the present study (observation extracted from CTX global mosaic). Two main areas are investigated with CRISM data: Area A in late winter to early spring ($L_\mathrm{S}\,183^\circ$ and $L_\mathrm{S}\,227^\circ$), and Area B in summer($L_\mathrm{S}\,318^\circ$), with temporal coverage constrained by observational availability.}
    \label{fig:surface+Sisyphi}
\end{figure*}

On Mars, distinctive downslope flow structures, known as "gullies" \citep{Malin2000}, have been observed and bear a striking resemblance to features carved by liquid water on Earth. These gullies have likely been forming since a few million years ago \citep{Reiss2004, Schon2009, DeHaas2015}, and are found predominantly in the mid-latitudes between 30° and 50° \citep{Balme2006, Noblet2024}. The distribution of these features shows a higher concentration in the Southern Hemisphere (Figure~\ref{fig:surface+Sisyphi} left panel, \citep{Harrison2015, Noblet2024}). Various morphological types of gullies have been identified, varying from linear to highly sinuous forms, and occurring in diverse settings, from rocky terrains such as crater walls and central peaks to surface sediments like sand dunes \citep{Auld2016a}. Most Martian gullies show three main morphologic features: an alcove (ranging from narrow to wide), where material is eroded in a dendritic pattern; an incised channel that transports the eroded material; and/or a fan-shaped apron at the base, where the material is ultimately deposited. While these features are frequently present together, some gully types could lack one component: linear gullies typically lack an apron, and some gullies are composed of an alcove and an apron only \citep{Auld2016a}. While most gullies are currently dormant, some remain active \citep{Dundas2010, Dundas2012, Dundas2022}, especially in the Southern Hemisphere. This contemporary activity appears to be closely linked to the presence of seasonal surface ice \citep{Dundas2015, Raack2015, Pasquon2023}. 

\FloatBarrier

Multiple hypotheses have been proposed to explain both the initial formation millions year ago and the contemporary activity of these structures. Initially, water was considered the main factor responsible for shaping and modifying gullies, based especially on analogies with Earth. Proposed mechanisms include the discharge of subsurface aquifer \citep{Malin2000} or water ice transitioning into a liquid state and flowing down the hillslope \citep{Costard2002, Lange2026}. Gully formation may thus have occurred hundred thousand years ago, when orbital parameters could have favoured water ice melting \citep{Clow1987, Costard2002, Williams2009}. However, the discovery of ongoing erosion of gully channels \citep{Diniega2010, Dundas2010, Dundas2015} suggests that other mechanisms may be at work, as the present-day occurrence of water ice melting or brine formation is considered an unlikely mechanism due to Mars' current climatic conditions \cite{Lange2026}. Other studies, such as \cite{Hoffman2002}, \cite{Ishii2006}, \cite{Diniega2010} or \cite{Dundas2015}, introduced and prioritized another key factor that could explain gullies' initial formation and/or recent modifications: CO\textsubscript{2} ice, formed from the deposition of the dominant constituent of Mars’ atmosphere. Various mechanisms have been proposed to explain the erosional processes at play in this case \citep{Bridges2001, Ishii2005, Cedillo-Flores2011, Vincendon2015, Pilorget2016, Khuller2021, Dickson2023, Dundas2025, Roelofs2025}, some of which require a specific physical state for CO\textsubscript{2} ice (granular, in blocks, translucent…). For instance, gully formation has been suggested to occur in association with geysers within a translucent slab of CO\textsubscript{2} ice \citep{Pilorget2016}, through the sliding of CO\textsubscript{2} ice blocks \citep{Diniega2013, Roelofs2025}, or via frost‐fluidization processes \citep{Dundas2025}. Alternatively, recent hypotheses also suggest a combination of H\textsubscript{2}O and CO\textsubscript{2} contributions. As an example, gullies may have initially formed with liquid water as the primary carving agent and current CO\textsubscript{2}-based modifications may only represent a second order process \citep{Dickson2023}. Other studies suggested that current modifications may be primarily driven by CO\textsubscript{2} ice, with some contribution of water ice for certain minor changes \citep{Vincendon2015, Khuller2021rev}. Finally, it has also been suggested that perennial subsurface water ice may act as a catalyst in the ongoing activity of gullies, by enhancing some CO\textsubscript{2}-driven mechanisms previously described \citep{Forget2024}.

Gullies on Mars have primarily been documented and studied using visible data, with instruments like MOC, HiRISE, HRSC and CTX \citep{Malin2000, Dundas2010, Dundas2012, Dundas2022, Harrison2015, Raack2015, Raack2020, Pasquon2019, Pasquon2023}. In contrast, infrared data —— both Near-IR \citep{Raack2015, Vincendon2015, Nunez2016, Andrieu2018, Allender2018, Khuller2021} and Thermal IR \citep{Raack2015, Khuller2021} —— have been used less frequently. Visible data allow for the observation of morphological processes in gullies, such as erosion and deposit formation, and can be used also to detect bright ice presence at high spatial sampling. Infrared spectroscopy can provides complementary insights: identification of ice composition (CO\textsubscript{2} and H\textsubscript{2}O), detection of ice contaminants, and determination of ice physical state (granular, translucent). In this study, our objective is to investigate Martian gullies through infrared spectroscopy, with two instruments: the Compact Reconnaissance Imaging Spectrometer for Mars (CRISM) \citep{Murchie2007} on NASA’s Mars Reconnaissance Orbiter, and the Observatoire pour la Minéralogie, l'Eau, les Glaces et l'Activité (OMEGA) \citep{Bibring2004} on ESA's Mars Express mission.
\\
Our research focuses on a high-latitude region in Mars Southern Hemisphere known as Sisyphi Cavi (68.5°S, 1.5°E), as shown in Figure~\ref{fig:surface+Sisyphi}. This site hosts active classic-type gullies \citep{Auld2016a}, particularly on equator-facing slopes as well as on South-East-facing slopes \citep{Raack2015, Raack2020, Pasquon2023}. Sisyphi Cavi is part of the few areas that lie outside the typical latitude band where most gullies are found, which makes it an interesting outlier to test possible gully formation/modification mechanisms. Additionally, the presence of extensive CO\textsubscript{2} geyser activity in this region further motivated its selection, as this process has been suggested to cause gully modifications \citep{Pilorget2016}.

\section{Methods}

\subsection{Overview}\

\begin{figure}[htbp]
    \centering
    \includegraphics[width=\columnwidth]{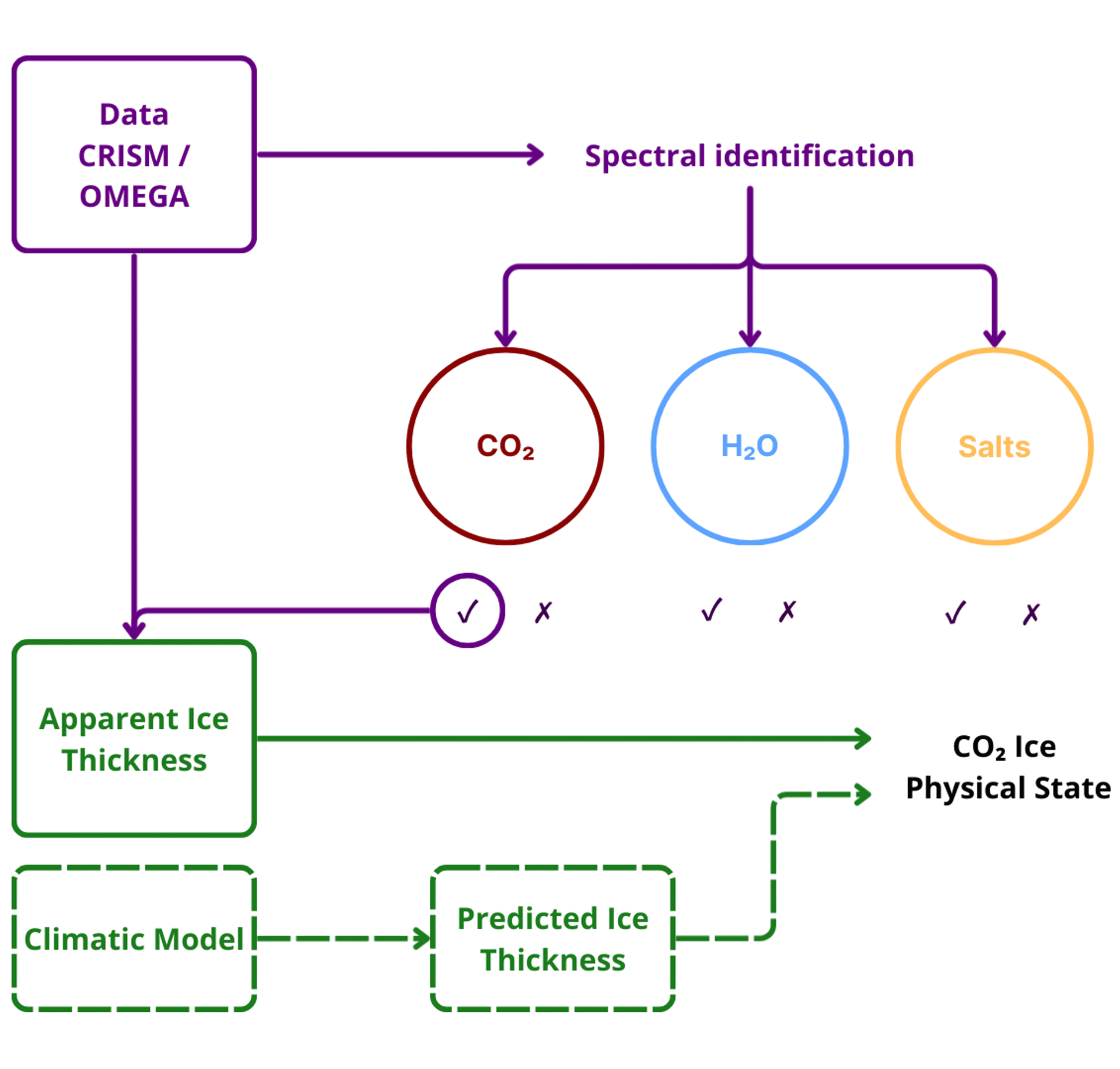}
    \caption{\textbf{General workflow of the method.} Using near-infrared spectral features, we first identify the presence of CO\textsubscript{2} ice, H\textsubscript{2}O ice, or salts on the studied site (Step 1, purple). If CO\textsubscript{2} ice is detected, a radiative transfer model is applied to estimate its thickness, which is then compared with outputs from the climatic model (Step 2, green) to estimate CO\textsubscript{2} ice physical state.}
    \label{fig:concept}
\end{figure}

We first provide an overview of the method used here to analyse near-IR data (illustrated in Figure~\ref{fig:concept}). We begin by analysing the spectral features observed in the gullied slopes of Sisyphi to constrain surface composition. We search for CO\textsubscript{2} ice and/or H\textsubscript{2}O ice presence during spring, when gully activity occurs, and for the identification of the site mineralogy in seasons without ice, as potential salts presence may promote liquid water activity \citep{Jouannic2019}.

\begin{figure}[htbp]
    \centering
    \includegraphics[width=\columnwidth]{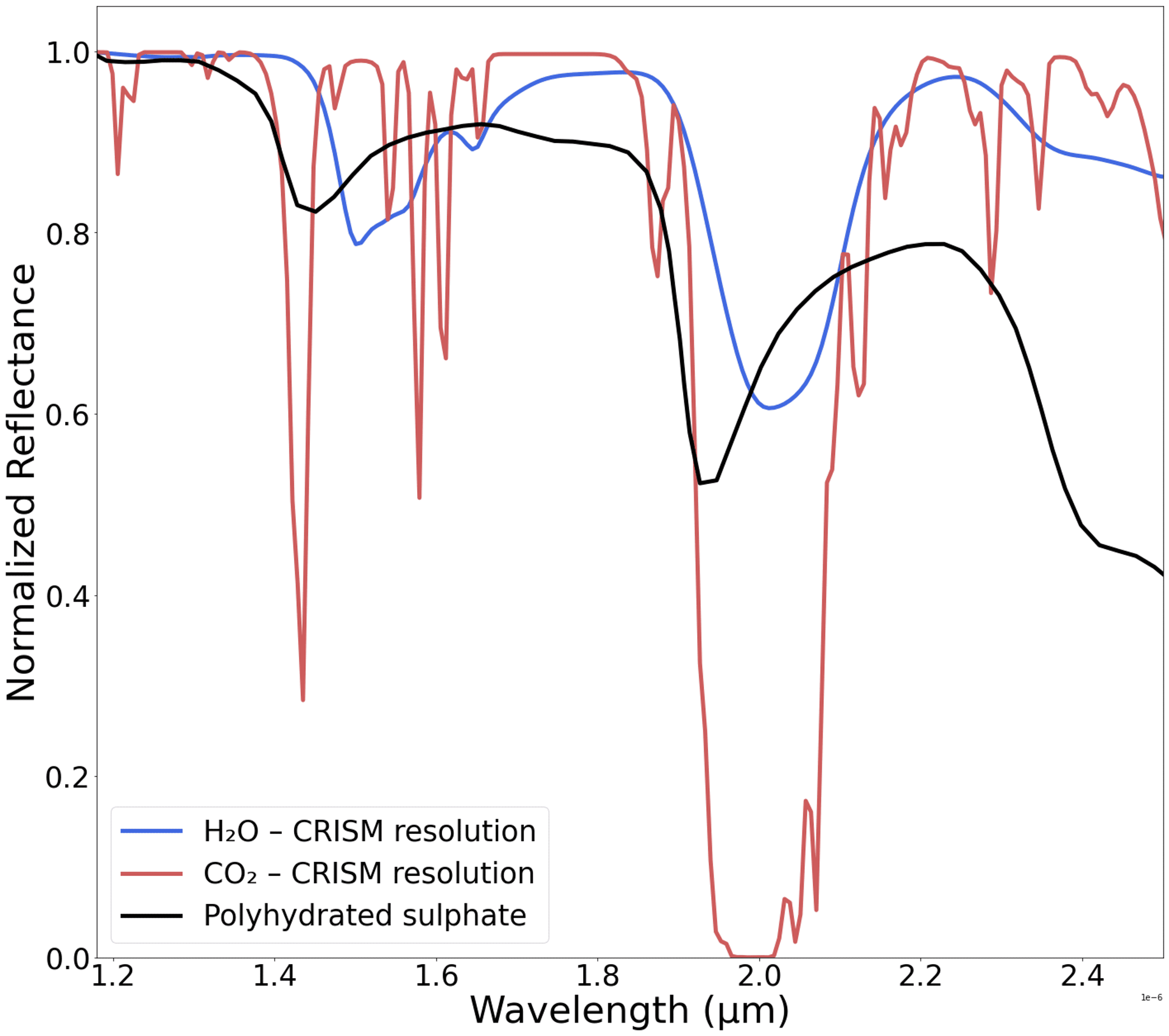}
    \caption{\textbf{Synthetic spectra of H\textsubscript{2}O ice, CO\textsubscript{2} ice, and a representative sulfate.} Synthetic CO\textsubscript{2} ice spectrum for a thick deposit ($L=1.10^{-1}m$, Equation~\ref{eq:synth}) in red, and synthetic H\textsubscript{2}O ice spectrum for a thin deposit ($L=5.10^{-5}m$, Equation~\ref{eq:synth}) in blue,  both derived using the methods described in Section 2.3.1. In black, a laboratory spectrum of a representative poly-hydrated sulfate, Epsomite (MgSO\textsubscript{4}·7H\textsubscript{2}O), is displayed \citep{Arvidson2005}. CO\textsubscript{2} and H\textsubscript{2}O ice spectra are shown at the CRISM spectral resolution of 6.55 nm. All spectra are normalized to 1 for better clarity.}
    \label{fig:IdSpec}
\end{figure}

Ice composition is found using spectral band identification \citep{Langevin2007, Carter2013} and/or full spectrum modelling (see Figure~\ref{fig:IdSpec}). Then, we focus on the properties of CO\textsubscript{2} ice, determining its physical state (granular or translucent) through the combination of the results derived from two modelling approaches. Radiative-transfer models are first used to interpret observed spectra and convert the optical path into an observed "apparent" ice thickness. Then, a 1D climatic model is used to simulate how much ice is present at the surface over time, which provides a "predicted" ice thickness. We then compare this apparent thickness to the one estimated by the climatic model: if the data and model align in trend and/or in values, the ice could be in a translucent state; if they do not, it suggests that the ice is more likely in a granular state.

\subsection{CRISM and OMEGA data}\

The CRISM dataset used in this study consists of three targeted observations on both areas A and B (Figure~\ref{fig:surface+Sisyphi}) : two Full Resolution Targeted (FRT) cubes (FRT00011935 and FRT00007D1E) acquired at $L_\mathrm{S}\,227^\circ$ and $L_\mathrm{S}\,318 ^\circ$, respectively, with a spatial resolution of $18~\pm~2$ m/pixel, and one Half Resolution Long (HRL00010475) cube acquired at $L_\mathrm{S}\,183^\circ$ with a spatial resolution of 36 m/pixel (Table~\ref{tab:observations}). Observations at $L_\mathrm{S}\,183^\circ$ and $L_\mathrm{S}\,227^\circ$ were the only two available during the ice season. The observation at $L_\mathrm{S}\,318 ^\circ$ was selected to estimate the presence of salts in the ice-free summer conditions. To ensure consistency and minimize noise across the dataset, CRISM observations during ice season were spatially averaged to a common footprint of 600 m/pixel (Table~\ref{tab:observations}). This spatial averaging helps mitigate potential biases caused by noise and avoid small-scale heterogeneities in spectra. Several corrections were applied to improve the reliability of the spectra. First, the raw reflectance of CRISM data $R_\mathrm{raw}(\lambda)$ has been corrected for atmospheric effects, with two independent corrections, one for gas, and one for dust particles (Table~\ref{tab:observations}). The gas correction corresponds to the default “volcano-scan” method provided by the CRISM analysis tool \citep{McGuire2009}. The dust particle correction is the one presented in \cite{Vincendon2007}. We have applied this correction with a dust optical depth of 0.2 +/- 0.1 according to measurements obtained over the polar cap in that season \citep{Vincendon2008}. These two corrections provide $R_\mathrm{atmo}(\lambda)$. Then, we apply the following additional corrections:

\begin{equation}
R_\mathrm{corr}(\lambda) = \frac{R_\mathrm{atmo}(\lambda)}{a(\lambda)}\times f_0
\label{eq:corr}
\end{equation}

where $a(\lambda)$ represents the continuum of the spectrum and $f_{0}$ a corrective factor for shadow effect. We correct by $a(\lambda)$ to account for the fact that our regions of interest correspond to cold slopes at the end of the polar night, mainly illuminated by light scattered by dust particles. This illumination contains a decreasing spectral slope (see e.g. \citep{Vincendon2008}) and this effect is not accounted for in the correction by \citep{Vincendon2007} which relies on a parallel-plane approximation (such an approach is designed for flat terrains and moderate solar zenith angles). We calculate $a(\lambda)$ using a polynomial function constrained by selected anchor wavelengths $\lambda_i$ free of absorption features: 1.15 µm, 1.37 µm and 2.21 µm. Additionally, since some slopes are entirely in geometric shadow, their calculated reflectance is artificially low. We correct for this by multiplying the reflectance by a factor $f_{0}$, derived from the ratio between the reflectance of nearby flat terrains and the observed spectrum, thereby normalizing shadowed spectra to a common continuum level. $f_{0}$ is close to 1 for the $L_\mathrm{S}\,227^\circ$ observation, but reaches 1.4 for some spectra of the $L_\mathrm{S}\,183^\circ$ observation.
The spectral-smile artefact, known to affect CRISM data \citep{Murchie2007} by shifting wavelengths near image edges, is also taken into account, but later on when spectra are modelled by radiative-transfer algorithms (see section 2.4).

\begin{table*}[hbt!]
\centering
\caption{\textbf{Summary of OMEGA and CRISM observations used in this study, with Solar Longitude, Martian Year, average pixel size and binning, spectral resolution and corrections applied.}}
\label{tab:data}
\begin{tabular}{|c|c|c|c|c|c|c|}
  \hline
   & \textbf{HRL00010475} & \textbf{FRT00011935} & \textbf{ORB1899\_2} & \textbf{ORB1921\_2} & \textbf{ORB1943\_2} & \textbf{FRT00007D1E} \\
  \hline
  \textbf{Solar Longitude (°)} & 183 & 227 & 245 & 249 & 253 & 318 \\
  \hline
  \textbf{Martian Year} & 29 & 29 & 27 & 27 & 27 & 28 \\
  \hline
  \makecell{\textbf{Average pixel size (m)} \\ \textbf{\textit{(with binning)}}} 
& \makecell{36 \\ \textit{(600)}} 
& \makecell{18\\ \textit{(600)}} 
& \makecell{3 200} 
& \makecell{2 900} 
& \makecell{2 600} 
& \makecell{18} \\
   \hline
  \textbf{Spectral resolution (nm)} & 6.55 & 6.55 & 13 & 13 & 13 & 6.55\\
   \hline
  \textbf{Corrections} & $^{a+b}$ & $^{a+b}$ & $^{a}$ & $^{a}$ & $^{a}$ & $^{a}$ \\
  \hline
\end{tabular}
\begin{flushleft}
\small
\textsuperscript{a} CO\textsubscript{2} + H\textsubscript{2}O gas, using the volcano scan method. \\
\textsuperscript{b} Atmospheric dust, calculated using the method described in \citep{Vincendon2007}.
\end{flushleft}
\label{tab:observations}
\end{table*}

OMEGA observations, with spatial resolutions ranging from 2.6 to 3.2 km/pixel, were added to complement CRISM by providing temporal coverage of surface ice during gully activity. Among the OMEGA cubes available over Sisyphi, three were retained: ORB1899\_2 ($L_\mathrm{S}\,245^\circ$), ORB1921\_2 ($L_\mathrm{S}\,249^\circ$), and ORB1943\_2 ($L_\mathrm{S}\,253^\circ$), as they sample the late spring to constrain the transition from icy to ice-free conditions. The areas of interest are covered only by a few pixels within the OMEGA observations. All OMEGA cubes went through standard atmospheric and thermal corrections following established processing protocols \citep{Langevin2007}. We summarise data information and procedures in Table~\ref{tab:data}.

\subsection{Ice spectral modelling}
\subsubsection{Beer-Lambert approach}\

We model the ice spectrum in order to extract the optical path of photons through the layer. To do so, we assume that the spectra observed correspond to a pure transparent ice, without inclusions. Then, we produced a spectral database of synthetic spectra $R_{\text{synth}}(\lambda)$ using Beer-Lambert’s law and reference optical constants \citep{Quirico2004CO2, Schmitt2004H2O}. Our library finally covers optical path lengths from $1.10^{-3}$ to 2 m with a step of $1.10^{-3}$. At this point, we take into account the spectral smile artefact \citep{Murchie2007} by adding a "free" step of $\pm~6.55 \text{nm}$, noted $\lambda'$, to improve band parameter retrievals:

\begin{equation}
R_{\text{synth}}(\lambda') = \exp\big(-k(\lambda')L\big)
\label{eq:synth}
\end{equation}

where $L$ is the optical path length in the ice layer and $k=\left(\frac{4\pi k_{im}}{\lambda'}\right)$, with $k_{im}$ the extinction coefficient. The spectra are then convolved to CRISM's spectral resolution (6.55 nm) by applying a Gaussian distribution on each point of the spectra, with the FWHM set to match the sampling of the CRISM data.
To complete the synthetic database, we introduce a neutral contaminant spectrum $C(\lambda)$, representing surface deposits, atmospheric dust, or embedded scatterers. A linear mixing is then performed between $R_{\text{synth}}(\lambda')$ and $C(\lambda') = d\lambda' + b$. The mixing coefficient is $x$, and the resulting synthetic reflectance is $R_{\text{synth'}}(\lambda')$:

\begin{equation}
R_{\text{synth'}}(\lambda') = (x~\times~R_{\text{synth}}(\lambda'))~+~(1-x)~\times~C(\lambda')
\label{eq:fit}
\end{equation}

where $x$ is the part of "pure" ice, and $1-x$ is the part of contaminants.

We then build a lookup table of possible $R_{\text{synth'}}$ as a function of the free parameters $L$, $\lambda'$, and $d$. The parameter $b$ is not treated as a free variable, as it is constrained by the continuum level of the spectrum being fitted.
 
\begin{figure}[t!]
    \centering
    \includegraphics[width=\columnwidth]{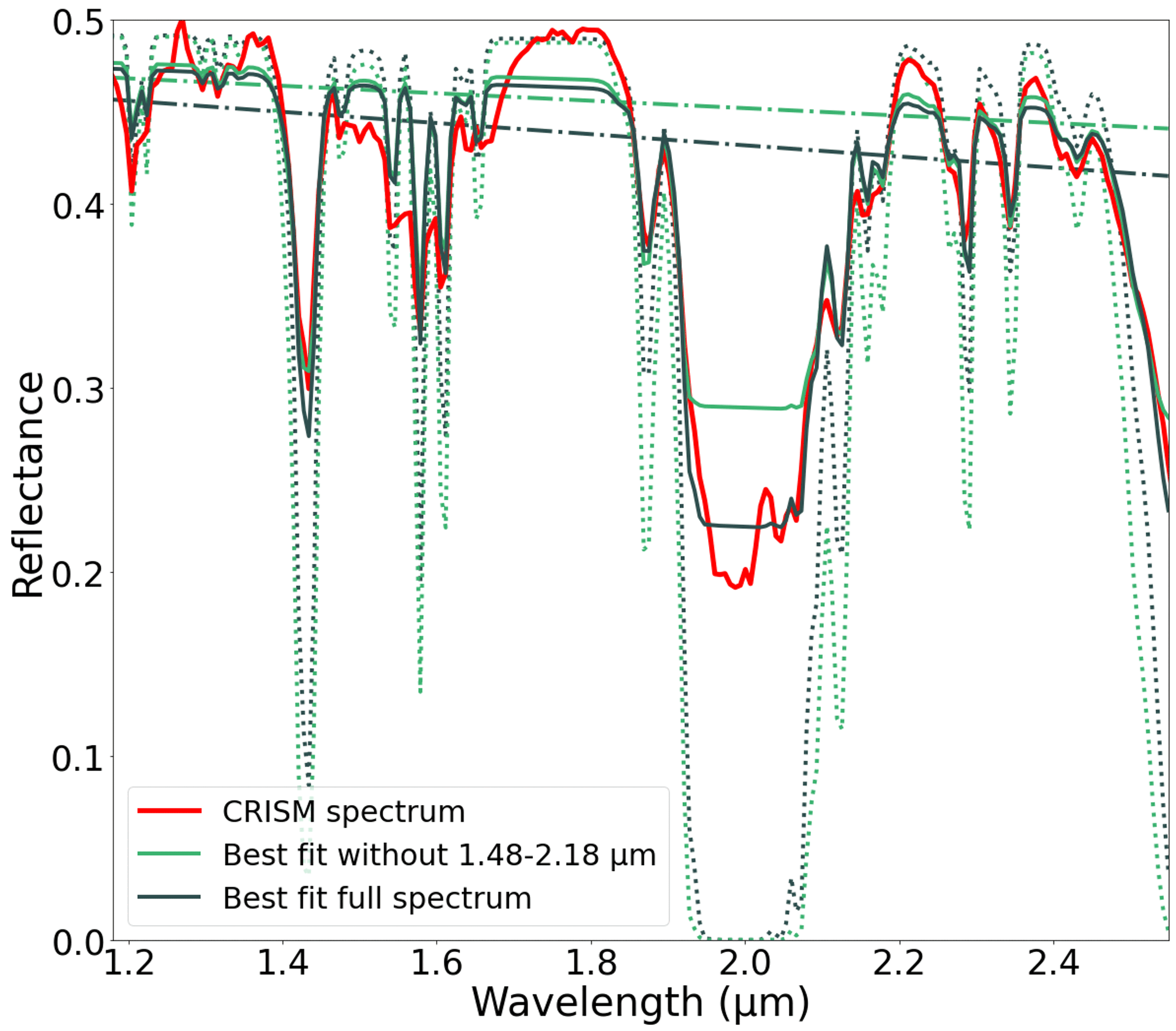}
    \caption{\textbf{Example of a Beer-Lambert fit of a CRISM CO\textsubscript{2} spectrum at $L_\mathrm{S}\,183^\circ$ for the N-facing slope (0$^\circ$).} In solid red line, corrected CRISM data; dotted line, pure CO\textsubscript{2} model; dash-dot line, best-fit contaminant; solid line, final fit. For green lines, the fit yields an optical path of 333 mm and $\approx 65\%$ contamination, applying an exclusion zone between 1.48–2.18 µm. For black lines, the fit gives an optical path of 181 mm and $\approx 50\%$ contamination without applying an exclusion zone.} 
    \label{fig:fitexemple}
\end{figure}

Finally, we perform a quadratic minimisation between $R_{\text{corr}}(\lambda)$ and $R_{\text{synth'}}(\lambda', L, d)$ over a spectral range composed of two intervals: 1.18 to 1.48~µm and 2.18 to 2.5~µm. The best output to this minimisation is $L_{\text{best}}$.
The lower bound at 1.18~µm is imposed by the absence of experimental data from optical constants used below this wavelength. 
The choice to exclude the 1.48–2.18~µm interval was guided by direct comparison of fits performed with and without this interval, as some spectral regions were better reproduced without (see Figure~\ref{fig:fitexemple}). This may notably be related to instrumental biases previously reported \citep{Wolff2019}. Spectral minimisation over the whole 1.18–2.5~µm range was however also conducted to assess potential systematic biases resulting from this choice (see section 3.3.1).

\begin{table*}[]
\caption{\textbf{Direct and scattered illumination conditions used in the modelling of the two icy CRISM observations (see Equation~\ref{eq:thick}).} The 8 columns correspond to 8 spectra extracted on icy slopes of both CRISM observations in area A (Figure~\ref{fig:surface+Sisyphi}) at 8 azimuthal angles. Solar zenith angles of direct/scattered ($i_{\text{d}}$ and $i_{\text{s}}$ respectively) light and fraction of direct/scattered ($F_{\text{d}}$ and $F_{\text{s}}$ respectively) light are indicated.}
\centering
\renewcommand{\arraystretch}{1.3}
\begin{tabular}{|c|c|c|c|c|c|c|c|c|c|}
\hline
\makecell{\textbf{Orientation}\\\textbf{Slope azimuth}} & 
& \makecell{\textbf{NNW}\\(337.5°)} &
\makecell{\textbf{N}\\(0°)} &
\makecell{\textbf{NNE}\\(22.5°)} &
\makecell{\textbf{NE}\\(45°)} &
\makecell{\textbf{NEE}\\(67.5°)} &
\makecell{\textbf{E}\\(90°)} &
\makecell{\textbf{SEE}\\(112.5°)} &
\makecell{\textbf{SE}\\(135°)} \\
\hline

\multirow{4}{*}{\makecell{\textbf{HRL}\\($L_\mathrm{S}\,183^\circ$)}} 
  & \textbf{$i_{\text{d}}$} & 71° & 76° & 79° & 84° & 88° & 90° & 91° & 90° \\ \cline{2-10}
  & \textbf{$F_{\text{d}}$} & 0.25 & 0.22 & 0.16 & 0.10 & 0.04 & 0.00 & 0.00 & 0.00 \\ \cline{2-10}
  & \textbf{$i_{\text{s}}$} & 55° & 60° & 60° & 70° & 70° & 70° & 70° & 70° \\ \cline{2-10}
  & \textbf{$F_{\text{s}}$} & 0.75 & 0.78 & 0.84 & 0.90 & 0.96 & 1.00 & 1.00 & 1.00 \\ \cline{2-10}
\hline

\multirow{4}{*}{\makecell{\textbf{FRT}\\($L_\mathrm{S}\,227^\circ$)}} 
  & \textbf{$i_{\text{d}}$} & 55° & 52° & 60° & 64° & 71° & 72° & 72° & 71° \\ \cline{2-10}
  & \textbf{$F_{\text{d}}$} & 0.46 & 0.48 & 0.38 & 0.35 & 0.26 & 0.25 & 0.25 & 0.26 \\ \cline{2-10}
  & \textbf{$i_{\text{s}}$} & 50° & 50° & 50° & 50° & 60° & 60° & 60° & 60° \\ \cline{2-10}
  & \textbf{$F_{\text{s}}$} & 0.54 & 0.52 & 0.62 & 0.65 & 0.74 & 0.75 & 0.75 & 0.74 \\ \cline{2-10}
\hline
\end{tabular}
\label{tab:diffusion}
\end{table*}

Once the optical path length ($L_{\text{fin}}$) is retrieved, it is converted into physical apparent ice thickness ($T_{\text{app}}$) using topographic data from MOLA (Mars Orbiter Laser Altimeter \citep{Smith2001}, onboard Mars Global Surveyor (MGS), NASA) combined with CRISM illumination geometry, accounting for both slope and internal refraction. Scattered illumination is accounted for by considering a statistical distribution of photon arrival directions at the surface \citep{Vincendon2009}. The effective mean incidence angles derived from this distribution shown in Table~\ref{tab:diffusion} are implemented to  represent the resulting illumination geometry within the ice layer. The apparent ice thickness $T_{\text{app}}$ is expressed as follows:

\begin{equation}
T_{\text{app}} = F_{d}\left( \frac{L_{fin}}{\frac{1}{\cos\alpha} + \frac{1}{\cos i'_{\text{d}}}} \right)+F_{s}\left( \frac{L_{fin}}{\frac{1}{\cos\alpha} + \frac{1}{\cos i'_{\text{s}}}} \right)
\label{eq:thick}
\end{equation}
\begin{equation*}
\text{with}~i' = \arcsin\left(\frac{\sin i}{n_{\text{CO}_2}}\right)
\end{equation*}

where $i_{\text{d}}$ is the local direct incidence angle at the surface, $i_{\text{s}}$ the mean scattered incidence angle, $\alpha$ the slope angle, $F_{d}$ the fraction of direct illumination, and $F_{s}$ the fraction of scattered illumination, as described in Table~\ref{tab:diffusion}. $i'$ is  the refracted angle inside the ice layer associated with a given $i$. Including this correction alter the retrieved thicknesses by only a few millimetres, indicating that the atmospheric scattered component have a minimal impact on the observed spectra. This limited effect is due to light refraction within the ice, as described by Snell’s law, which reduces the effective contribution of high incidence angle incoming photons. The real part of the refractive index of CO\textsubscript{2} ice, $n_{\text{CO}_2}$, is assumed to be 1.4, following the value reported by \cite{Quirico2004CO2}. 
We estimate the uncertainties ($\delta$) based on twice the minimum value of the quadratic criterion, yielding $L_{\text{final}}=L_{\text{best}} \pm\delta\text{, and}~\delta=2q_{min}$ where $q_{min}$ is the quadratic minimum found for $L_{best}$. The confidence interval for $T_{\text{app}}$ is derived from the $L$ values at the lower and upper bounds of the $L_{\text{fin}}$ confidence interval in Equation~\ref{eq:thick}, yielding $T_{\text{app, min}}$ and $T_{\text{app, max}}$.

\subsubsection{\cite{Andrieu2018} approach}\

As a complementary approach, a second modelling method is applied to provide an alternative estimation of ice properties from observed spectra. The method previously validated by \cite{Andrieu2018} is used to simulate spectra of layers of ice with more complex properties \citep{Andrieu2015}, with a different inversion procedure based on Bayesian statistics \citep{Mosegaard1995, Andrieu2016}. In further detail, this method propagates the uncertainties from the data to the model parameters and provide the distribution of all possible solutions. The model used laboratory optical constants for CO\textsubscript{2} and water ice \citep{Schmitt1998}, while Martian dust optical properties are estimated from CRISM measurements on defrosted areas. It includes translucent CO\textsubscript{2} ice slab thickness, impurities (water ice inclusions, dust), surface roughness, and linear mixtures with H\textsubscript{2}O ice and defrosted terrains to account for sub-pixel heterogeneities \citep{Andrieu2018}.

\subsection{Climatic model predictions}\

Finally, we perform a calculation of ice thickness predicted by a climatic model for Sisyphi Cavi. As explained in the overview (Figure~\ref{fig:concept}), this predicted thickness will be compared to the apparent thickness $T_{\text{app}}$ previously derived from observation to evaluate CO\textsubscript{2} ice physical state (granular or translucent). If the thickness retrieved from spectroscopy is equal to the climatic simulation, then the ice must be translucent to the bottom. If the thickness retrieved from spectroscopy is lower than the climatic simulations, then the ice could be either translucent in the upper part or granular (as proposed in \cite{Andrieu2018}).
We use outputs from a 1D version of a Global Climatic Model \citep{Forget1999} previously adapted to water ice and compared to near-IR data by \cite{Vincendon2010}. This model simulates the seasonal evolution of CO\textsubscript{2} ice and H\textsubscript{2}O ice deposits based on key parameters such as thermal inertia, slope magnitude and orientation, geographic coordinates, solar longitude (L\textsubscript{s}), permafrost depth, as well as the albedo of both the surface and the ice layer. Model outputs include the mass of CO\textsubscript{2} ice that accumulate and then sublimate at the surface. This mass is converted to a "predicted" ice thickness using a density of $1.62~\mathrm{g.cm^{-3}}$ \citep{Mangan2017}, consistent with compact crystalline ice for 140--150~K (\citep{Raack2015}). Since ice density depends on deposition temperature, lower temperatures could increase it up to $\sim1.68~\mathrm{g\,cm^{-3}}$. However, density variations would only cause minor shifts in absolute thickness values without affecting the overall trends in "predicted" thickness.
\\
In the next sections, we have tuned some parameters of the climatic model to obtain various plausible predictions of ice thicknesses to be compared to observations, following the approach described in previous studies using this model \citep{Vincendon2010, Vincendon2015}:
\begin{itemize}[--]
    \item permafrost depth (between 0.02 and 0.2~m)
    \item CO\textsubscript{2} ice albedo (between 0.2 and 0.8)
    \item surface albedo (between 0.2 and 0.4)
\end{itemize}
Other parameters of the model were set to fixed values consistent with the areas we analysed in Sisyphi Cavi, such as slopes angles or thermal inertia (200~$\mathrm{J\,m^{-2}\,K^{-1}\,s^{-1/2}}$ according to \cite{Christensen1992}).

We must notice that this early version of the climatic model has been upgraded since then \citep{Lange2023}. However, we do not aim at using the climatic model to constrain with observations the parameters of the Sisyphi area. The climatic model is used to provide general trends about ice thickness as a function of slope azimuth angles or solar longitude. A missing heat source can e.g. be replaced by tuning the other parameters.

\begin{figure*}[!b]
    \centering
    \includegraphics[width=\textwidth]{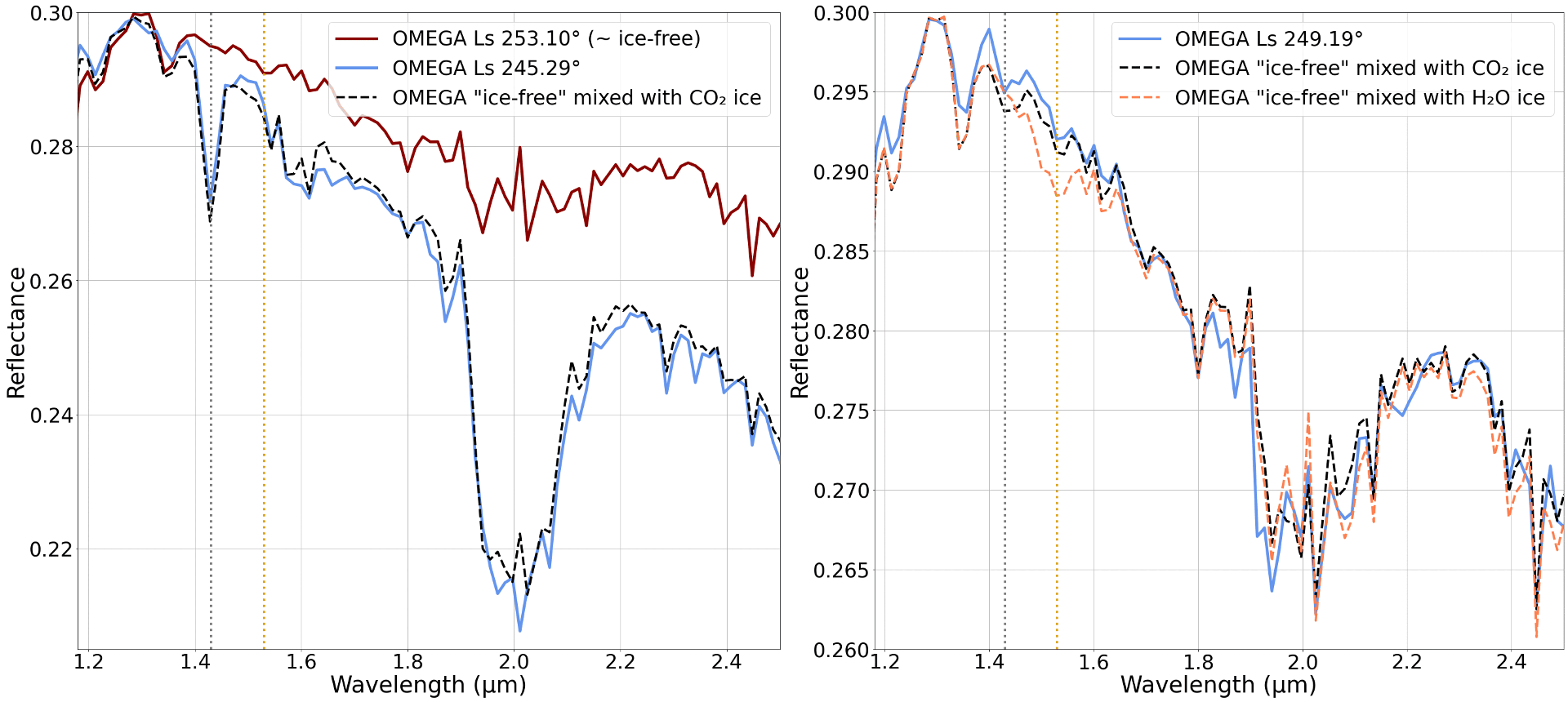}
    \caption{\textbf{Identification of ice in the final stage of the frost presence at Sisyphi Cavi by OMEGA.} Left panel: OMEGA spectra acquired at $L_\mathrm{S}\,245^\circ$ (blue) and $L_\mathrm{S}\,253^\circ$ (dark red). Ice signatures are no longer detected for observation at  $L_\mathrm{S}\,253^\circ$ (dark red), making this spectrum an ice-free reference to model other observations (both panels). In black dotted line, best Beer-Lambert model fit for $L_\mathrm{S}\,245^\circ$ observation spectrum, modelled with CO\textsubscript{2} ice for $L = 78.10^{-3} mm$ (Equation~\ref{eq:synth}) and $x = 0.15$ (Equation~\ref{eq:fit}) to confirm CO\textsubscript{2} ice presence. Right panel: observation at $L_\mathrm{S}\,249^\circ$ compared with two Beer–Lambert models: one using only CO\textsubscript{2} ice ($L = 5.10^{-3} m$ (Equation~\ref{eq:synth}) and $x = 0.05$ (Equation~\ref{eq:fit})), and one using only H\textsubscript{2}O ice ($L = 5.10^{-5}m$ (Equation~\ref{eq:synth}) and $x = 0.05$ (Equation~\ref{eq:fit})). Both ices can reproduce the strong 2~µm band, but only the synthetic CO\textsubscript{2} ice spectrum matches the 1.4–1.6~µm range. This region indeed contains a faint CO\textsubscript{2} ice 1.43~µm feature but no detectable 1.5~µm water ice feature. The CO\textsubscript{2} 1.43~µm band (vertical grey dotted line) and the H\textsubscript{2}O 1.5~µm band (vertical orange dotted line) are indicated in both panels.}
    \label{fig:OMEGA}
\end{figure*}

\section{Results}

\subsection{Ice presence}\

The CRISM observations of area A (Figure~\ref{fig:surface+Sisyphi}) acquired during the ice season were analysed to assess ice composition. Extracted spectra show several diagnostic features of CO\textsubscript{2} ice, and are well reproduced by mixtures of CO\textsubscript{2} ice with a spectrally neutral component (Figure~\ref{fig:fitexemple}). Faint H\textsubscript{2}O ice features may also be present within the CO\textsubscript{2}-dominated spectra, notably in the early spring observation at $L_S\,183^\circ$ (Figure~\ref{fig:fitexemple}), as suggested by previous studies \citep{Langevin2007, Raack2015}.

To extend the analysis to later seasonal stages, during the final steps of seasonal ice sublimation, three OMEGA observations are analysed. In the OMEGA data, a clear CO\textsubscript{2} ice signature is observed at $L_S\,245^\circ$ (Figure~\ref{fig:OMEGA}, left panel), although noticeably shallower than in earlier CRISM data. The $L_S\,249^\circ$ spectra show a faint, broad feature near 2.0 µm, potentially compatible with either CO\textsubscript{2} or H\textsubscript{2}O ice (Figure~\ref{fig:OMEGA}, right panel). However, spectral fitting over the entire range indicates that this feature can be entirely explained by small residual amounts of CO\textsubscript{2} ice. Indeed, the absorption band of H\textsubscript{2}O ice at 1.5 µm is absent, whereas a weak 1.43 µm CO\textsubscript{2} band is detected. The observation at $L_S\,253^\circ$ no longer show any detectable ice features. A faint broad and spatially uniform 2.0 µm feature persists, but without any correlation with cold slopes. Water ice is thus not detected in any of the three OMEGA observations, and CO\textsubscript{2} ice is detected up to $L_S\,249^\circ$, but is no longer present in the subsequent observation at $L_S\,253^\circ$. We should however note that some ice may be present at the surface or in the near-surface without being detected by near-IR spectroscopy, notably if ice is almost not illuminated, covered by an optically thick material or too optically thin to be observable by orbital spectrometers \citep{Valantinas2024}.

\subsection{Salts}\

\FloatBarrier
\begin{figure*}[!b]
    \centering
    \includegraphics[width=\textwidth]{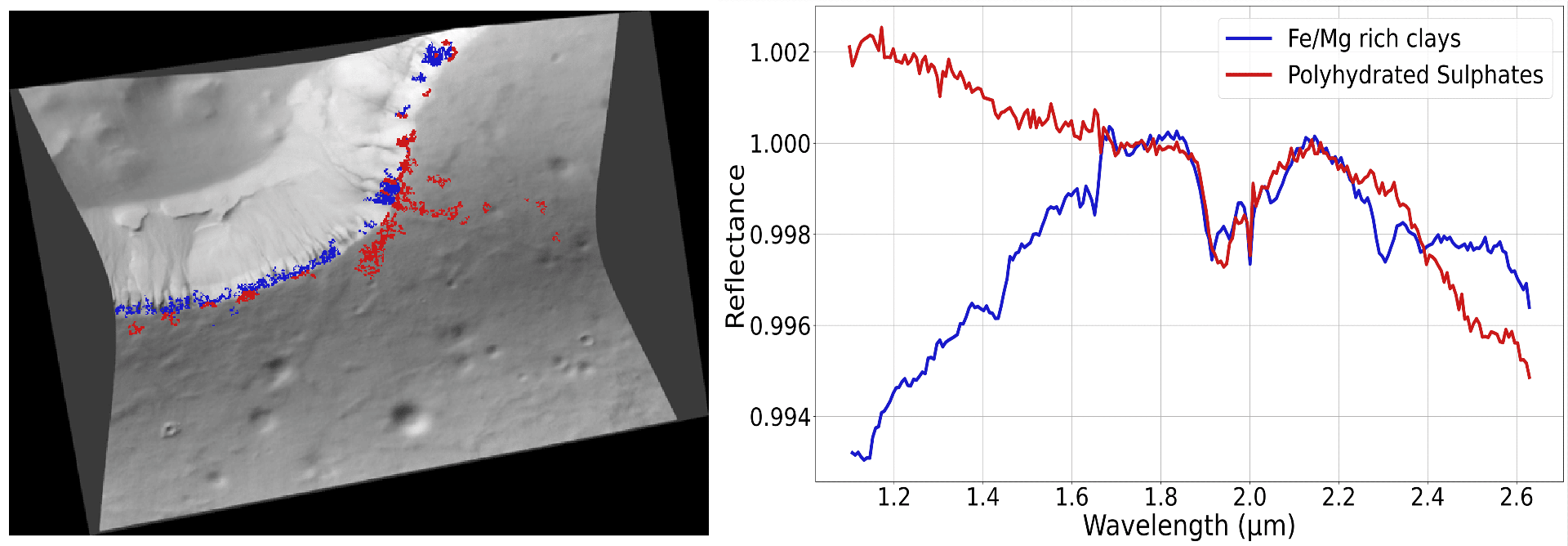}
    \caption{\textbf{Detection of sulfates and clays in the Sisyphi Cavi region.} CRISM observation FRT00007D1E (area B, Figure~\ref{fig:surface+Sisyphi}) at $L_\mathrm{S}\,318^\circ$ is used. Left panel: spatial distribution of poly-hydrated sulfates (red) and Fe/Mg-rich clays (blue) detections. Right panel: mean spectra extracted from the main sulfate-rich spot and from regions with clay detections.}
    \label{fig:detectionsalts}
\end{figure*}

Using CRISM observation acquired at $L_\mathrm{S}\,318^\circ$, we investigate the presence of salt and, additionally, other hydrated minerals, in the Sisyphi Cavi region (following the method used in \cite{Jouannic2019}). Spectral analysis reveal absorption features at 1.9 µm and 2.5 µm diagnostic of poly-hydrated sulfates, and at 2.3 µm indicative of Fe/Mg clays (Figure~\ref{fig:detectionsalts}, right panel). Sulfate detections are preferably distributed horizontally across the central part of the observation, extending over the upper part of the crater walls and adjacent plateau, but are absent from lower and middle parts of crater walls where the gully channels and aprons are situated (Figure~\ref{fig:detectionsalts}, left panel). The very weak band depths and restricted spatial distribution suggest that these detections are near the CRISM detection limit. Fe/Mg clays detections are less frequent and mainly detected in the upper alcove regions, where erosion appears to have exposed subsurface clay-bearing layers (Figure~\ref{fig:detectionsalts}, left panel).

\FloatBarrier
\begin{figure*}[!h]
    \centering
    \includegraphics[width=\textwidth]{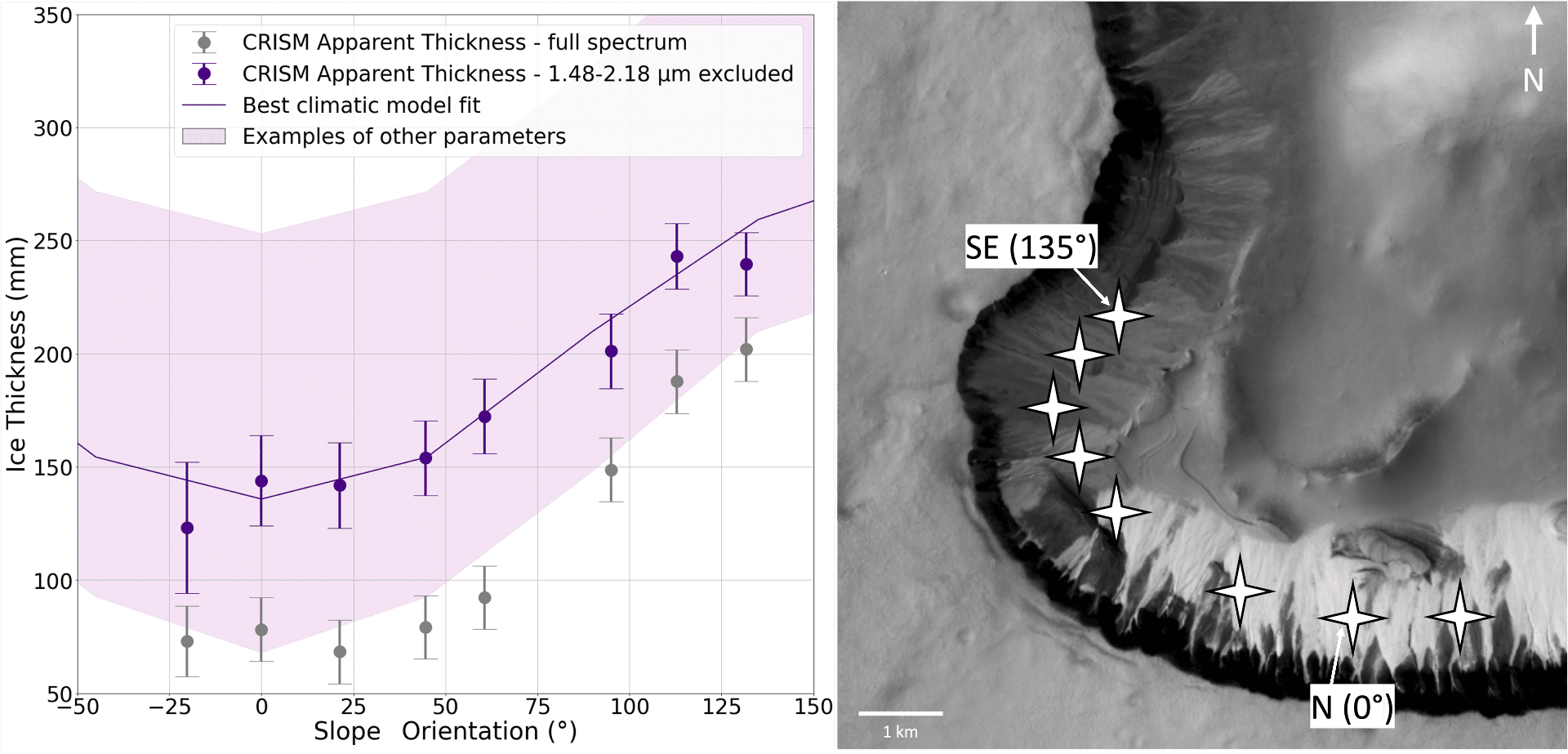}
    \caption{\textbf{Azimuthal distribution of ice thickness derived from CRISM observation compared to ice thickness predicted by the climatic model ($L_\mathrm{S}~183^\circ$).} Left panel: purple dots are the ice thickness retrieved from CRISM-derived optical paths using the Beer-Lambert modelling method, excluding the 1.48–2.18 µm spectral region. Grey dots: same retrievals but without excluding any spectral region (full spectrum, as shown in Figure~\ref{fig:fitexemple}). Purple line: predicted thickness from the climatic model, with the fit that reproduce best our data shown as a dark purple line and alternative models represented by the lighter purple area. The best-fitting model uses a CO\textsubscript{2} ice albedo of 0.5, a permafrost at 0.02 m, and a ground albedo of 0.2. Right panel: extracted observation from CTX global mosaic, showing the locations of extracted spectra (white stars) for context. N (0°) and SE (135°) orientations are indicated for clarity, corresponding respectively to "warmer" and "colder" slopes regions.}
    \label{fig:slopeori}
\end{figure*}

\subsection{CO\textsubscript{2} ice physical state}\

In this section, we compare apparent ice thickness (derived from CRISM data) to predicted ice thickness (obtained with the climatic model). First, we study the behaviour depending on slope azimuth orientation. Then we focus on the seasonal variations.

\subsubsection{Slope Orientation}\

We first perform this comparison with the Beer-Lambert approach at a single L\textsubscript{s} for various slope orientations (Figure~\ref{fig:slopeori}). This representation allows us to test the expected spatial distribution of ice: thicker deposits should occur on colder slopes (SE-facing, $135^\circ$), while thinner deposits are expected on warmer slopes (N-facing, $0^\circ$). A direct comparison with the climatic model then indicates whether the apparent ice thickness derived from CRISM is consistent with the climatic modelled values.

\FloatBarrier
\begin{figure*}[htbp]
    \centering
    \includegraphics[width=\textwidth]{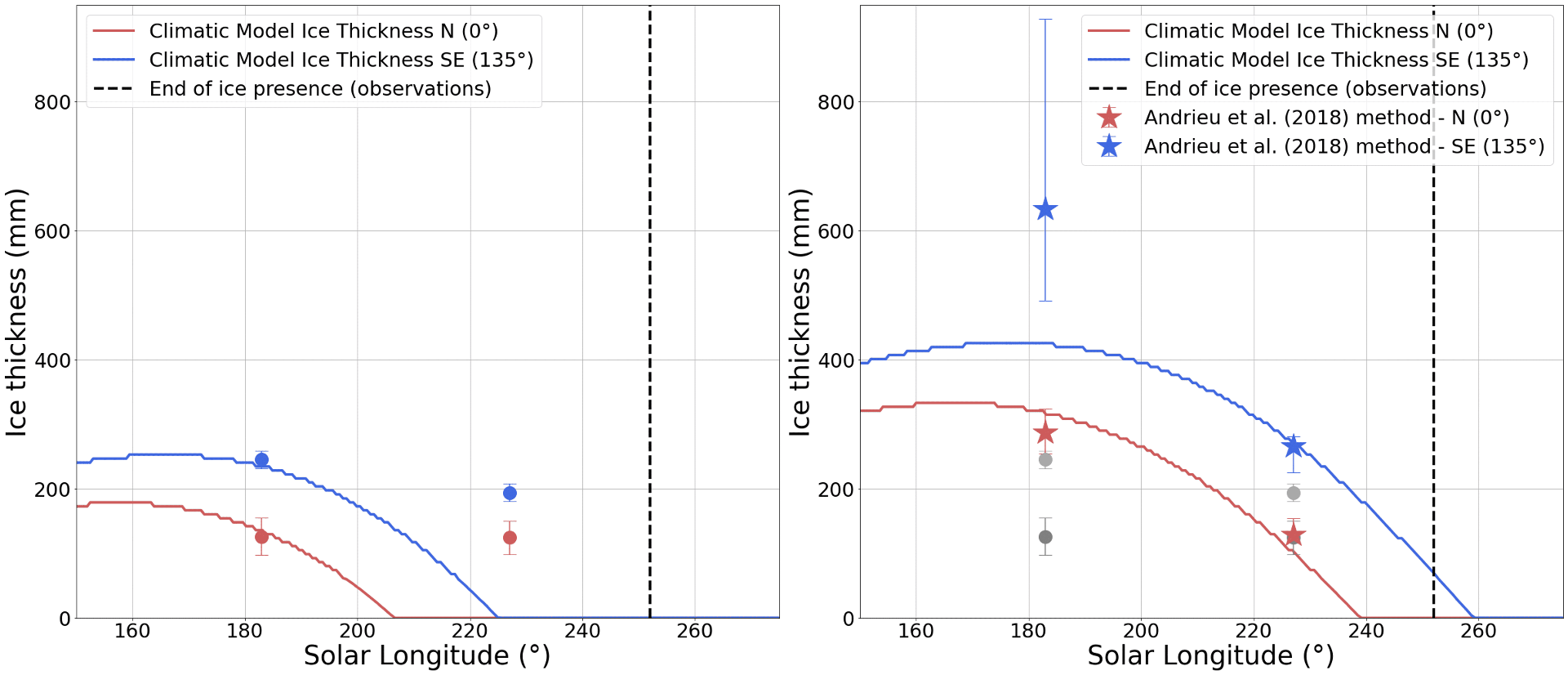}
     \caption{\textbf{Comparison between CRISM apparent ice thicknesses and climatic model predictions as a function of solar longitude, for $L_\mathrm{S}\,183^\circ$ and $L_\mathrm{S}\,227^\circ$.} Each panel corresponds to a different modelling of CRISM data: the left panel is the Beer-Lambert method, and the right panel is used as a comparison, with \cite{Andrieu2018} approach. Dots indicate CRISM apparent ice thickness derived with the Beer-Lambert method, stars  are \cite{Andrieu2018} model's outputs. In blue, SE (135°) oriented slopes (cold slopes). In red, N (0°) oriented slopes (warm slopes). Right panel, in grey: data from the left panel are repeated for comparison. The dashed black line shows the end of ice presence from OMEGA data. }
    \label{fig:CRISMLS}
\end{figure*}

In Figure~\ref{fig:slopeori}, we focus on the observation at $L_\mathrm{S}\,183^\circ$. The area covers slopes with orientations ranging from NNW-facing ($-25^\circ$) to SE-facing ($135^\circ$), providing a sufficiently wide range to capture spatial variations in ice distribution across the Sisyphi Cavi region. Eight representative slope orientations are selected, for which apparent ice thickness is calculated. A best-fit climatic model is identified, corresponding to the solution that most closely reproduces the apparent thicknesses. In this case, both the values and the overall trend agree well with the climatic model: retrieved thicknesses ranges from $\approx{120~mm}$ to $\approx{240~mm}$, with larger amounts of ice on the colder SE-facing slopes and reduced ice on the warmer N-facing slopes. An envelope of additional model solutions further shows that the observed trend remains robust, even if we perform the analysis of the CRISM data over the whole 1.18–2.5~µm spectral range (see section 2.3.1 and  Figure~\ref{fig:fitexemple}).

\subsubsection{Solar Longitude}\

In addition, we apply the same procedure, this time comparing the thickness as a function of solar longitude (Figure~\ref{fig:CRISMLS}) for two L\textsubscript{S}. The objective here is to evaluate the expected decreasing trend associated with sublimation during the late ice season, i.e., thinner ice at $L_\mathrm{S}\,227^\circ$ compared to $L_\mathrm{S}\,183^\circ$ as spring progresses. In this representation, we only show two representative slope orientations (N-facing at 0°E, and SEE-facing at 113°E), but both at $L_\mathrm{S}\,183^\circ$ and $L_\mathrm{S}\,227^\circ$. We first compare in the left panel apparent thickness to the predicted thickness corresponding to the "best-fit" at $L_\mathrm{S}\,183^\circ$ (Figure~\ref{fig:slopeori}). We can see that this "best-fit" predicts an early sublimation of ice for $L_\mathrm{S}\,210^\circ$ - $L_\mathrm{S}\,225^\circ$ that does not fit the data any more, neither the CRISM observation at $L_\mathrm{S}\,227^\circ$ nor the OMEGA data at $L_\mathrm{S}\,245^\circ$ - $L_\mathrm{S}\,249^\circ$ (see Figure~\ref{fig:OMEGA}).

\begin{figure}[htbp]
    \centering
    \includegraphics[width=\columnwidth]{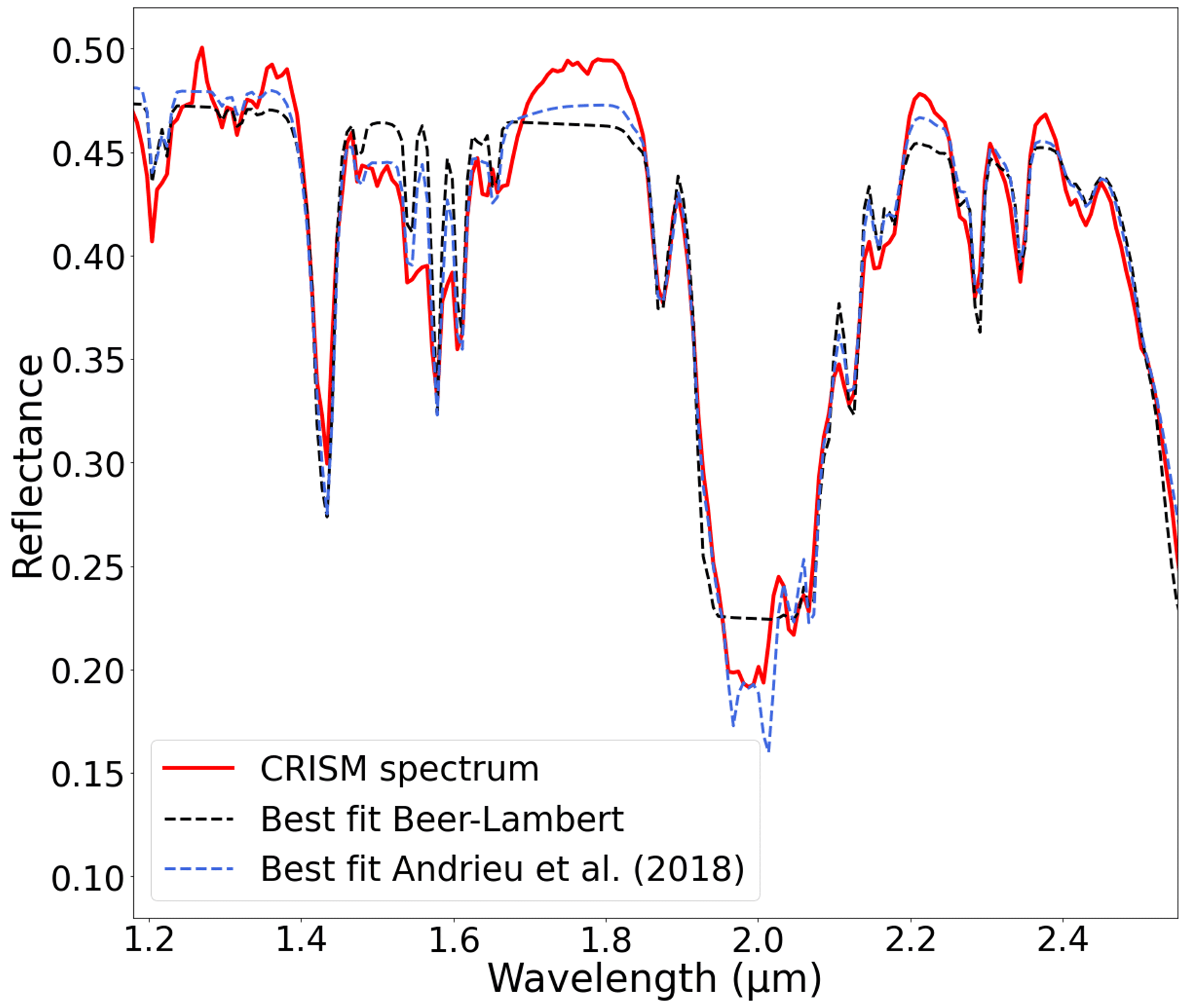}
     \caption{\textbf{Comparison between the Beer-Lambert modelling and the \cite{Andrieu2018} method, for a N-oriented slope at $L_\mathrm{S}\,183^\circ$.} The CRISM spectrum is shown as a red solid line. The Beer--Lambert fitting method is represented by a black dashed line, yielding an ice thickness of $126 \pm 29$ mm. The \cite{Andrieu2018} method is depicted as a blue dashed line, providing an ice thickness of $288 \pm [32, 36]$ mm.}
    \label{fig:BeerLambertVSAndrieu}
\end{figure}

\begin{table*}[!b]
\centering
\caption{\textbf{Summary of surface activity observed at the Sisyphi Cavi site.} Observations start around $L_\mathrm{S}\,160^\circ$, after the polar night.}
\resizebox{\textwidth}{!}{
\begin{tabular}{|p{2.2cm}|p{4.6cm}|p{3.0cm}|p{4.8cm}|p{0.7cm}|}
\hline
\textbf{Process} &
\textbf{Description} &
\makecell[l]{\textbf{Formation Timing} \\ \textbf{(Solar Longitude, Ls)}} &
\textbf{Slope Orientation} &
\textbf{Ref.} \\
\hline

\textbf{Dark spots} &
Dark spots above seasonal ice (interpreted as products of CO\textsubscript{2} geysers) &
\makecell[l]{$L_\mathrm{S}\,<160-180^\circ{}^{*}$ \\ $L_\mathrm{S}\,<160^\circ{}^{*}$} &
\makecell[l]{E and SE-facing \\ N-facing slopes} &
1, 2, 3 \\
\hline

\textbf{Dark flows} &
Dry granular downslope flows overlying CO\textsubscript{2} ice &
\makecell[l]{$L_\mathrm{S}\,215-226^\circ$, MY29 \\ $L_\mathrm{S}\,232-252^\circ$, MY33 \\ $L_\mathrm{S}\,226-247^\circ$, MY29} &
\makecell[l]{N-facing slope (large event) \\ SEE-facing slope (large event) \\ SE-facing slope (smaller event)} &
1, 2 \\
\hline

\textbf{Gully \ modifications} &
Erosional and depositional changes of the ice-free underlying terrain &
\makecell[l]{$L_\mathrm{S}\,215-247^\circ$, MY29 \\ $L_\mathrm{S}\,232-252^\circ$, MY33 \\ $L_\mathrm{S}\,226-247^\circ$, MY29} &
\makecell[l]{Erosion: N-facing \\ Deposition: E to SE-\\facing}&
1, 2 \\
\hline

\end{tabular}}
\label{tab:activity_summary}
\begin{flushleft}
\small
\textsuperscript{$^*$} between $L_\mathrm{S}\,180^\circ$ and $L_\mathrm{S}\,230^\circ$, dark spots persist on E and SE-facing slopes ; between $L_\mathrm{S}\,160^\circ$ and $L_\mathrm{S}\,230^\circ$, dark spots fade on N-facing slopes \\
\textsuperscript{1} \cite{Raack2015}
\textsuperscript{2} \cite{Raack2020}
\textsuperscript{3} \cite{Pasquon2023}
\end{flushleft}
\end{table*}

In order to better understand this difference between apparent and predicted thicknesses, we perform a similar comparison changing two elements. First, the apparent ice thickness derived from CRISM data is estimated using the \cite{Andrieu2018} approach instead of the Beer–Lambert one (comparison between the two methods is showed on Figure~\ref{fig:BeerLambertVSAndrieu}). Second, the climatic model parameters are adjusted to enable a longer ice stability, more consistent with the end of ice presence inferred from OMEGA observations $L_\mathrm{S}\,245^\circ$ - $L_\mathrm{S}\,249^\circ$. For that, the following parameters were used: CO\textsubscript{2} ice albedo of 0.5, permafrost at 0.07 m and albedo ground of 0.3. The results of this alternative comparison are shown in the right panel of Figure~\ref{fig:CRISMLS}.
The \cite{Andrieu2018} approach produces a stronger decrease of the apparent thickness with L\textsubscript{s}, which is more consistent with the recalculated climatic model predictions.

\section{Discussion}\

We will now discuss possible gully modification mechanisms considering these observational constraints. We first summarize the types and timing of activity identified at Sisyphi Cavi (section 4.1), then assess mechanisms involving water (section 4.2), followed by those involving CO\textsubscript{2} (section 4.3). Finally, we explore the implications for other gullies (section 4.4).

\subsection{Overview of activity timing}\

\begin{figure*}[htbp]
    \centering
    \includegraphics[width=\textwidth]{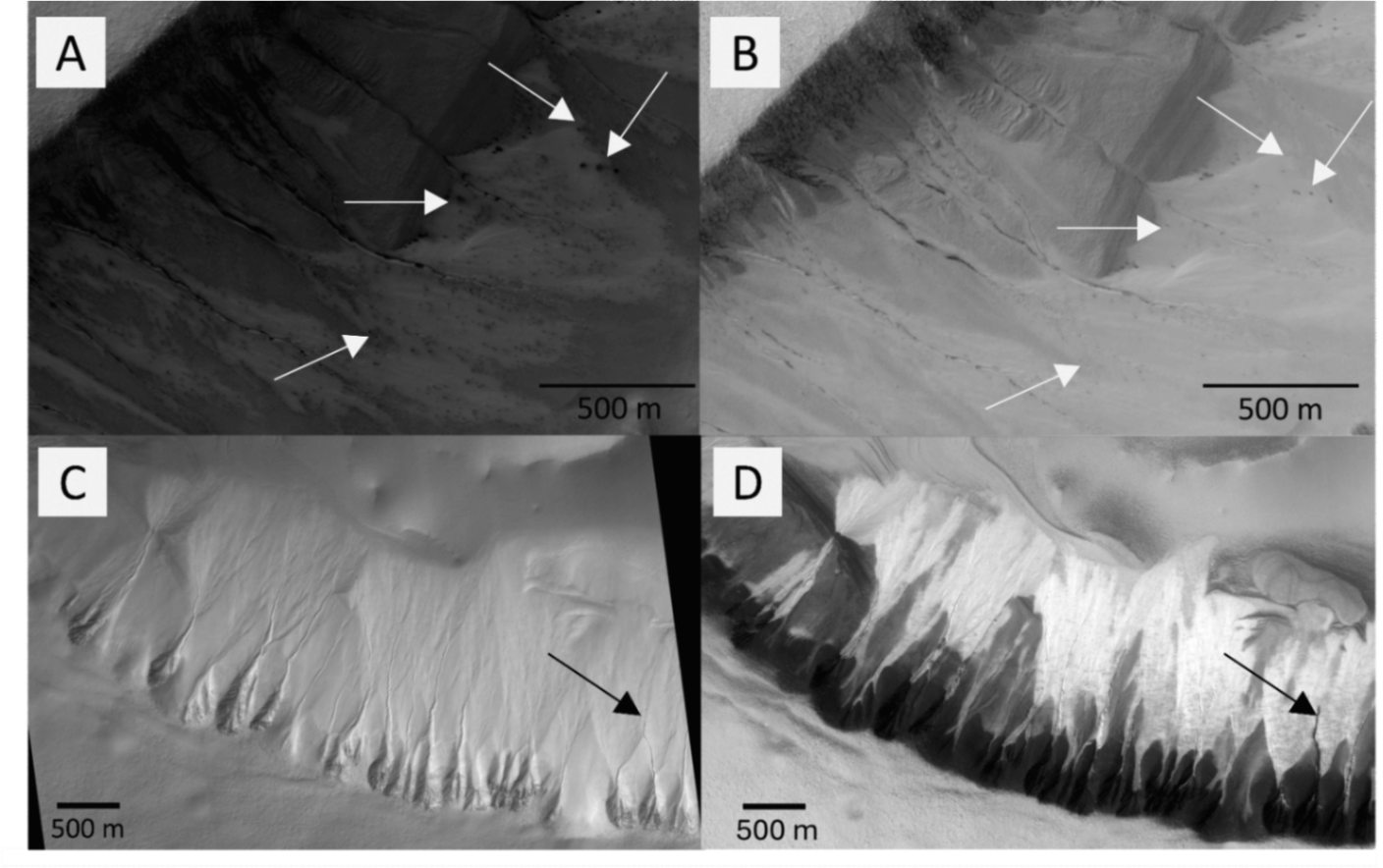}
    \caption{\textbf{HiRISE observations for \(L_\mathrm{S}\,183^\circ\) (ESP\_011396, panels A and C) and \(L_\mathrm{S}\,227^\circ\) (ESP\_012332, panels B and D), showing dark spots and dark flows.}
Panels A and B are the SE-facing slopes (135°), the coldest ones.
Panels C and D are the N-facing slopes (0°), the warmest ones.
Freshly formed dark spots can be seen at \(L_\mathrm{S}\,183^\circ\) on the SE-slope (A, white arrows).
Most of these dark spots have faded or disappeared at \(L_\mathrm{S}\,227^\circ\), and no new spots formed between the two observations (B, white arrows).
A major dark flow occurred between the two observations—marked by the black arrow in panels C (before) and D (after)—and eroded the corresponding gully, as reported by \cite{Raack2015} (see Table~\ref{tab:activity_summary}).
North is up on all panels.}
    \label{fig:DarkSpots}
\end{figure*}

We establish a summary table of activity timing and surface modification types at Sisyphi (Table~\ref{tab:activity_summary} and Figure~\ref{fig:DarkSpots}), built on previous investigations \citep{Raack2015, Raack2020, Pasquon2023} that primarily relied on HiRISE visible data. 

Several types of activity have been reported in the Sisyphi Cavi region, spanning from late winter to mid-spring. For clarity, these processes can be divided into three categories: (i) dark spots, (ii) dark flows, and (iii) gully modifications. 
\\

\textit{(i) Dark spots.} These features are generally interpreted as surface expressions of CO\textsubscript{2} geyser activity. In this process, solar radiation penetrates the translucent seasonal ice layer and warms the underlying regolith. This heating triggers CO\textsubscript{2} sublimation, which progressively builds up pressure beneath the ice. The pressure is then released through a geyser that also ejects small amounts of ice and regolith (Figure~\ref{fig:DarkSpots}, panels A and B). These spots are frequently observed at southern polar latitudes during late winter \citep{Kieffer2006}.
\\

\textit{(ii) Dark flows.} They correspond to downslope flows with a lower albedo compared to the icy slope (Figure~\ref{fig:DarkSpots}, panels C and D). Dark flows can be associated or not with dark spots and gully channels \citep{Dundas2012, Raack2020}. Some authors have suggested that different sub-types or categories of dark flows may exist on icy slopes, depending on their association with dark spots timing, introducing the concept of Recurrent Diffusing Flow \citep{Pasquon2016}.
\\

\textit{(iii) Gully modifications.} This category corresponds to morphological changes (erosion, deposition) within gullies, observed in ice-free conditions. Modifications are usually persistent once they have occurred. 
\\

Table~\ref{tab:activity_summary} shows that dark spots typically form early in the season, and can persist up to $L_S\,230^\circ$, mainly on E and SE-facing slopes (coldest slopes). In contrast, dark flows and gully activity occur later, coinciding with the end of ice season and the end of CO\textsubscript{2} ice sublimation. Dark flows appear on both N and SE-facing slopes, and gully activity is observed on N-facing slopes for erosion and E to SE-facing slopes for deposition \citep{Raack2015, Raack2020}. This timing illustrates a two-stage seasonal progression of surface activity, also modulated by slope orientation (occurring earlier on warmer slopes and later on colder ones), with dark spots forming near the equinox and gully processes occurring as the ice starts to disappear. Gully activity happened simultaneously with dark flows, or just after their occurrence.

\FloatBarrier

\subsection{The Role of Water}\

Previous studies have suggested that water may be the main initial carving agent of gullies, during past periods \citep{Malin2000, Costard2002, Dickson2023}, when atmospheric density was more favourable for water at low latitudes. It has also been suggested that water ice may be involved in some activity currently observed at gully sites, either as ice, in the same way as CO\textsubscript{2} but in smaller proportion, or via an ephemeral transition to liquid water (e.g., \citep{Vincendon2015, Jouannic2019, Khuller2021rev}). In the Sisyphi Cavi region, some water ice is probably present as inclusions within the CO\textsubscript{2}-dominated ice (see section 3). However, we do not observe times or places when/where water ice is present alone at the surface. Such a situation may occur at the mid-latitudes, where water ice can remain exposed for several tens of degrees of L\textsubscript{S} after CO\textsubscript{2} ice sublimation \citep{Vincendon2015}. This behaviour is not observed at Sisyphi Cavi: the three OMEGA observations covering the final $10^\circ$ of L\textsubscript{S} of the ice season show no detectable water-ice signature,  thus agreeing with the conclusions of \cite{Langevin2007}.
Summer data display a very faint, broad 2.0~µm absorption, no longer associated with any topographic feature as would be expected for residual water ice in cold traps. This pattern, consistent with features reported by \cite{Poulet2008} and more recently discussed by \cite{Stcherbinine2021} and \cite{Barraud2024}, more likely reflects regional sulfate signatures.

CRISM data do reveal the presence of sulfates in the area B (Figure~\ref{fig:surface+Sisyphi}) under ice-free conditions, but these detections are concentrated on the plateau rather than on the crater walls or at or at its floor. The last two would have indicated a possible link with gullies, either as deposits formed during sublimation or as an agent locally lowering the frost point where gullies occur. This distribution suggests that sulfates are more likely related to regional mineralogical processes than to active gully modification \citep{Carter2016}. This interpretation is reinforced by the fact that Fe/Mg clays are found near the upper depression edges, in zones affected by gully head erosion, implying that erosion may have simply exposed pre-existing layers rather than formed new alteration features (see Figure~\ref{fig:detectionsalts}, left panel).
In addition, the sulfate detections are near the CRISM sensitivity threshold (see Figure~\ref{fig:detectionsalts}, right panel) and appear detectable only in the central third of the scene, a pattern that could be influenced by photometric biases such as emergence angle variations. Finally, we observe very faint signature of sulfates with < 0.3\% band depth at 1.9 µm. This suggests that sulfates are probably present in minor amounts at the surface in the Sisyphi area. For comparison, previously identified localised high concentration of sulfates at high latitudes comes with very strong near-IR absorption features, of a few tens of \% \citep{Langevin2005}. The identified spectral signatures of sulfates at Sisyphi thus probably correspond to a few weight \% or less. This would be comparable to values inferred at the Phoenix landing site, that is situated at a similar latitude as Sisyphi but in the Northern Hemisphere \citep{Kounaves2010, Toner2014}. Such amount appears too small to lower the freezing point of water, compared to the several tens of percent suggested by previous studies \citep{Hennings2013}.

Taken together, these observations provide no evidence for water ice or liquid-water related processes currently active at Sisyphi Cavi.

\subsection{CO\textsubscript{2} ice properties and role}\

\begin{figure*}[h!]
    \centering
    \includegraphics[width=\textwidth]{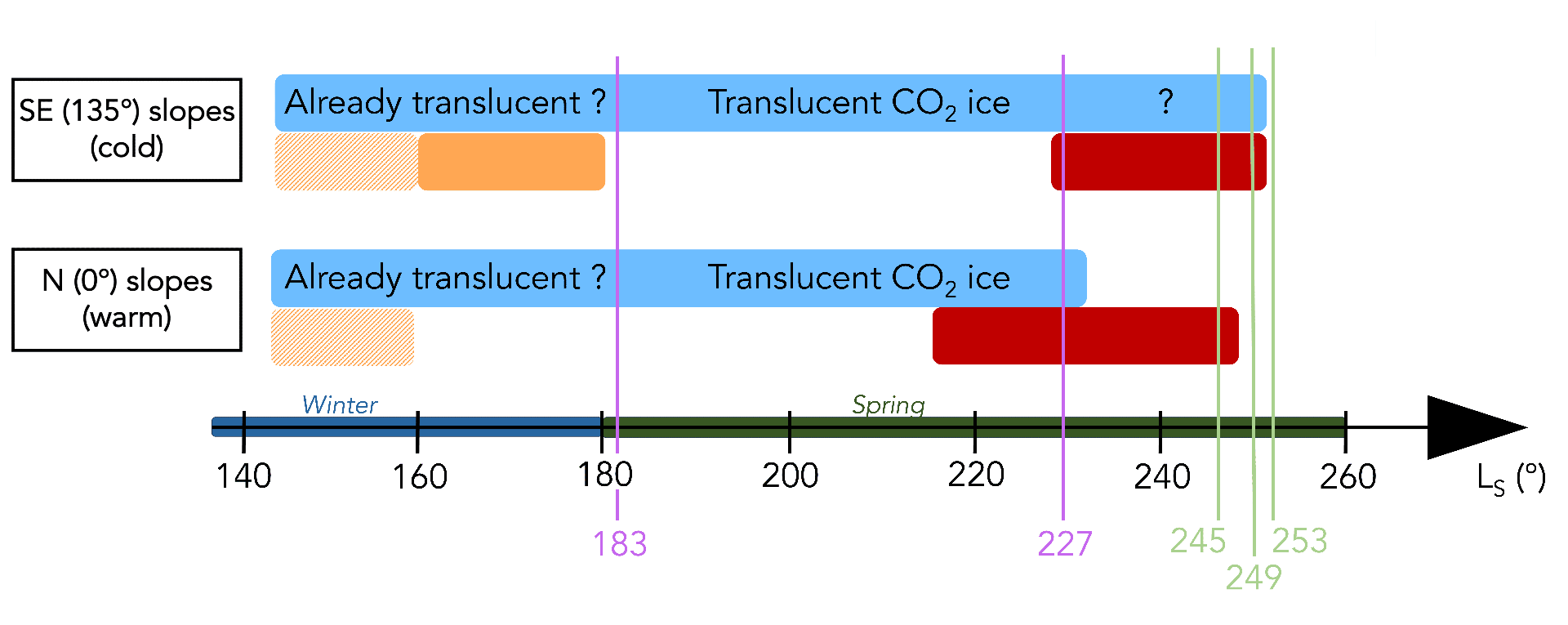}
    \caption{\textbf{Timeline comparing activity type and ice properties.} Summary timeline showing the timing of: ice presence, dark spot formation, and gully activity (erosion and deposition) for two main slope orientations : SE-facing slopes ("cold" slopes) and N-facing slopes ("warm" slopes). In blue, presence and physical state of CO\textsubscript{2} ice as determined in this study, complemented by observations from \cite{Raack2015, Raack2020} and \cite{Pasquon2023} for ice presence. In red, timing of gully activity, and in orange dark spot activity, both extracted from the same references; hatched orange denotes inferred activity for which only later remnants were observed. CRISM (purple) and OMEGA (light green) observations used in this study are also shown for clarity.}
    \label{fig:conclu}
\end{figure*}

On the other hand, the presence of CO\textsubscript{2} ice is well correlated with gully activity at Sisyphi, as already noted in previous studies \citep{Raack2015, Raack2020, Pasquon2023}. We illustrate this correlation in Figure~\ref{fig:conclu}.
Results presented in section 3.3 show that the observed optical path within the ice is sensitive to the amount of ice. This is observed as a function of azimuth (Figure~\ref{fig:slopeori}) and solar longitude when using the \cite{Andrieu2018} approach (Figure~\ref{fig:CRISMLS}). In both cases, relative variations agree well with predictions from the climatic model. This means that when the model predicts a thicker ice layer either at given L\textsubscript{s} or on a given slope orientation, the corresponding observed ice spectrum does show a longer optical path in the ice. These results show that ice is sufficiently translucent to enable photons to probe the whole ice layer and thus to reach the underlying surface. In addition, modelled ice thicknesses from the optical path are also broadly consistent in absolute values with expectations from the climatic model, being in the tens of centimetres range. This also supports the fact that most of the ice layer is probed by photons.

The lower ice thicknesses retrieved by the Beer-Lambert approach likely result from methodological differences: the \cite{Andrieu2018} approach accounts for contaminants, and the retrieved parameters from the fit indicate that the ice contains diffusive impurities (water ice inclusions, dust) that reduce the optical path within the ice.

Additionally, the mismatch between both models is reduced at $L_\mathrm{S}\,227^\circ$ compared to $L_\mathrm{S}\,183^\circ$, which suggests that the amount of contaminant may decrease as sublimation progresses. This is consistent with the mechanism  proposed by \cite{Pommerol2011JGR}, involving the formation and sinking of a dusty crust subsequently mobilised by surface winds as the ice sublimates.

\FloatBarrier

However, even if CO\textsubscript{2} ice forms a translucent slab, this physical state may not be to be a key driver of the CO\textsubscript{2} ice mechanism that modifies the gullies at Sisyphi. We observe that CO\textsubscript{2} ice is already translucent in the earliest CRISM observation at $L_\mathrm{S}\,183^\circ$, yet gully changes occur only around or after $L_\mathrm{S}\,227^\circ$ (Figure~\ref{fig:conclu}). For most of the period during which translucent CO\textsubscript{2} ice is present, no gully activity is detected. In addition, translucent ice is central to the CO\textsubscript{2} geyser mechanism \citep{Pilorget2016}, but the observed timing of gully modification at Sisyphi ($L_\mathrm{S}\,>215^\circ$) is incompatible with the season of geyser activity inferred from dark spots, which occurs much earlier ($L_\mathrm{S}\,<180^\circ$). Previous work has also shown that the total CO\textsubscript{2} ice condensing at low‐latitude active sites may be too low to sustain geyser-driven processes \citep{Vincendon2015}.
Taken together, these observations indicate that the presence of translucent CO\textsubscript{2} ice and geyser activity appears likely coincidental regarding gully activity. The timing and physical constraints instead point toward sublimation-driven granular flow processes, which remain consistent with the persistence of only a thin CO\textsubscript{2} layer in late spring \citep{Cedillo-Flores2011, Roelofs2024}.

\subsection{Implications for other gully sites}\

We observed that gully activity at Sisyphi Cavi occurs during phases of decreasing ice thickness (>$L_\mathrm{S}\,215^\circ$ for north-facing slopes and >$L_\mathrm{S}\,226^\circ$ for South East-facing slopes), after the seasonal maximum has been reached, and near the end of seasonal ice stability. This behaviour suggests that a critical factor controlling activity may be related to the transient conditions associated with ice retreat and final-stage sublimation.

This result could also be relevant for lower latitudes sites with smaller ice thicknesses, in agreement with some cases reported in \cite{Vincendon2015}. However, caution is required when extrapolating observations at Sisyphi Cavi to gullies in general. Our interpretations may not be directly applicable to linear or dune gullies, which exhibit markedly different morphologies and are associated with distinct substrates and sedimentary contexts \citep{Roelofs2024, Roelofs2025}. The processes discussed here for Sisyphi Cavi may therefore be more relevant to gully systems with comparable morphologies (i.e. "classic gullies", \citep{Auld2016a}) and formation settings.

\section{Conclusion}\

We investigated the formation and modification mechanisms of Martian gullies by focusing on an active site located at high southern latitude (68°S, 1°E):~Sisyphi Cavi. The processes driving this activity remain incompletely constrained, with several competing hypotheses involving CO\textsubscript{2} ice , H\textsubscript{2}O ice, or a combination of both. Using infrared data analysis (CRISM, OMEGA), the objective of this study was to characterize the properties, composition, and structure of the ice present during gully activity.

Our main findings are summarized as follows:

\begin{itemize}
    \item Based on previous studies, gully modification at Sisyphi Cavi is reported at the end of the ice season, between $L_\mathrm{S}\,215^\circ$ and $L_\mathrm{S}\,252^\circ$. Erosion and deposition within the gullies are spatially associated with the formation of dark flows.
    \item During this active period, we do not detect water ice in the OMEGA observations acquired at the end of the ice season, while CO\textsubscript{2} ice is still present and then sublimates ($L_\mathrm{S}\,245^\circ$, $L_\mathrm{S}\,249^\circ$ and $L_\mathrm{S}\,253^\circ$).
    \item We identified salts in the area, but their presence shows no spatial correlation with the gullies. They are mainly located on the plateau and near depression rims, likely exposed by erosion, and are also observed at the regional scale.
    \item CO\textsubscript{2} ice observed by CRISM at two different times ($L_\mathrm{S}\,183^\circ$ and $L_\mathrm{S}\,227^\circ$) exhibits a translucent behaviour in both cases. This suggests that CO\textsubscript{2} ice remains in a translucent state during most of the ice season, although its state during the very last stages (after $L_\mathrm{S}\,227^\circ$) remains uncertain.
    \item When comparing the reported timing of gully erosion with the occurrence of dark spots, well established indicators of CO\textsubscript{2} geyser activity, we found a clear temporal mismatch. Dark spots form much earlier in the season (from $L_\mathrm{S}\,160^\circ$ to $L_\mathrm{S}\,180^\circ$ depending on slope orientation) while gully activity occurs much later, near the disappearance of surface ice.
\end{itemize}

In summary, Sisyphi Cavi hosts active gullies on N and SE-facing slopes, with activity occurring at the end of the ice season ($L_\mathrm{S}\,215^\circ$–$252^\circ$). Our results do not support water-driven mechanisms, as no H\textsubscript{2}O ice is detected alone at the end of the ice season, and no spatial correlation is found between salts and gullies. CO\textsubscript{2} ice, by contrast, is present and remains translucent throughout the whole ice season.
Gully activity, however, is confined to the final stages, when the ice layer is thinner, and does not coincide with CO\textsubscript{2} geyser activity, which occurs earlier ($L_\mathrm{S}\,160^\circ$–$180^\circ$).
Taken together, these observations suggest that neither H\textsubscript{2}O ice nor CO\textsubscript{2} geysers alone can fully explain the gully activity at Sisyphi Cavi. Instead, the data favour CO\textsubscript{2}-ice-driven fluidization or avalanche processes that can operate during the late stages of seasonal ice sublimation.This conclusion may also apply to certain gullies located at lower latitudes.

\section*{Acknowledgements}
The Authors acknowledge support from the Centre National d’Études Spatiales (CNES).

\printcredits

\bibliographystyle{cas-model2-names}
% \bibliographystyle{spmpsci}
\bibliography{Mars_Biblio_final}

\end{document}